\documentclass[sigconf]{acmart}
\usepackage{graphicx} 
\usepackage{natbib}
\usepackage{subcaption}  
\usepackage{tabularx}
\acmConference[Conference]{34th ACM SIGSOFT International Symposium on Software Testing and Analysis}{September 2025}{xxxx, yyyy}
\usepackage{multirow}
\usepackage{algorithm}
\usepackage{algorithmicx}
\usepackage{algpseudocode}
\usepackage{setspace} %
\usepackage{caption}
\usepackage[utf8]{inputenc}
\usepackage[T2A]{fontenc}
\usepackage[russian,english]{babel}

\usepackage[many]{tcolorbox}  
\def\BibTeX{{\rm B\kern-.05em{\sc i\kern-.025em b}\kern-.08em
    T\kern-.1667em\lower.7ex\hbox{E}\kern-.125emX}}
\usepackage{tcolorbox}
\definecolor{main}{HTML}{5989cf}    
\definecolor{sub}{HTML}{cde4ff} 

\newtcolorbox{dashedbox}{
   colback = sub,
    enhanced,
    boxrule = 0.5pt, 
    colframe = white, 
    borderline = {.5pt}{0pt}{main, dashed}, 
     left=2.5pt, 
    right=2.5pt, 
    top=2pt, 
    bottom=2pt, 
    before skip=5pt, 
   after skip=5pt  
}
\newtcolorbox{boxJ}{
    sharpish corners, 
    colback = sub, 
    colframe = main, 
    boxrule = 0pt, 
    toprule = 4.5pt, 
    enhanced,
    fuzzy shadow = {0pt}{-2pt}{-0.5pt}{0.5pt}{black!35} 
    ,
     left=2.5pt, 
    right=2.5pt, 
     top=2pt, 
    bottom=2pt, 
    before skip=5pt, 
   after skip=5pt  
}

\newtcolorbox{boxH}{
    colback = gray!20,  
    colframe = gray!20, 
    boxrule = 0pt, 
    leftrule = 6pt, 
    left=2.5pt, 
    right=2.5pt, 
    top=2pt, 
    bottom=2pt, 
    before skip=5pt, 
    after skip=5pt  
}
\setcopyright{none} 
\renewcommand\footnotetextcopyrightpermission[1]{}

\begin{document}
\title{Learning-Based Testing for Deep Learning: Enhancing Model Robustness with Adversarial Input Prioritization}
\author{Sheikh Md. Mushfiqur Rahman}
    \affiliation{%
        \institution{University of Tennessee}
        \city{Knoxville}
        \state{TN}
        \country{USA}
}
\email{srahma14@vols.utk.edu}

\author{Nasir U. Eisty}
\affiliation{%
	   \institution{University of Tennessee}
	   \city{Knoxville}
	   \state{TN}
	   \country{USA}
}
\email{neisty@utk.edu}
\date{September 2025}

\begin{abstract}
\textbf{\textit{Context:}}
Deep Neural Networks (DNNs) are increasingly deployed in critical applications, where resilience against adversarial inputs is paramount. However, whether coverage-based or confidence-based, existing test prioritization methods often fail to efficiently identify the most fault-revealing inputs, limiting their practical effectiveness.
\textbf{\textit{Aims:}}
This project aims to enhance fault detection and model robustness in DNNs by integrating Learning-Based Testing (LBT) with hypothesis and mutation testing to efficiently prioritize adversarial test cases.
\textbf{\textit{Methods:}}
Our method selects a subset of adversarial inputs with a high likelihood of exposing model faults, without relying on architecture-specific characteristics or formal verification, making it adaptable across diverse DNNs.
\textbf{\textit{Results:}}
Our results demonstrate that the proposed LBT method consistently surpasses baseline approaches in prioritizing fault-revealing inputs and accelerating fault detection. By efficiently organizing test permutations, it uncovers all potential faults significantly faster across various datasets, model architectures, and adversarial attack techniques.
\textbf{\textit{Conclusion:}}
Beyond improving fault detection, our method preserves input diversity and provides effective guidance for model retraining, further enhancing robustness. These advantages establish our approach as a powerful and practical solution for adversarial test prioritization in real-world DNN applications.
\end{abstract}

\maketitle
\section{Introduction}
Data-driven software built on DNN models forms the backbone of various research fields, including autonomous driving~\cite{bojarski2016end}, image classification~\cite{he2016deep}, and speech recognition~\cite{xiong2016achieving}. DNN models comprise numerous interconnected neurons~\cite{xie2022npc,gerasimou2020importance}, resulting in highly complex internal parameters. Consequently, training and optimizing these models is a labor-intensive and time-consuming process that demands domain expertise~\cite{gao2022adaptive, shen2020multiple, wang2021prioritizing}. Despite the substantial computational cost, unresolved quality and reliability issues in DNNs can have severe consequences when confronted with adversarial examples (AEs), such as accidents involving self-driving cars~\cite{davies2019tesla, ziegler2016google}, potentially hindering the adoption of deep learning in safety-critical domains like medical diagnosis~\cite{zhang2020towards}. 

The software engineering community addresses this challenge through adversarial testing and verification methods. Testing DNN-based software fundamentally differs from traditional software testing, as it relies on data-driven programming rather than manually coded business logic. \citet{pei2017deepxplore} introduced Neuron Coverage as a testing criterion for DNNs, which paved the way for various testing metrics to evaluate DNN models~\cite{ma2018deepgauge}. Furthermore, researchers have explored adversarial sample generation through techniques such as white-box testing~\cite{pei2017deepxplore} and black-box testing~\cite{tian2018deeptest}. Another approach involves formal verification methods, such as SMT solving~\cite{sharma2021mlcheck}, to ensure that a DNN adheres to specific robustness properties. 

Nonetheless, these methods often entail high computational costs and are applicable only to a limited range of DNNs and properties. Additionally, executing DNN models demands significant computational resources, restricting the number of tests that can be performed~\cite{zolfagharian2024smarla}. To ensure that DNN adversarial testing remains both practical and cost-efficient, it is essential to select a smaller subset of test inputs with strong fault-revealing capabilities. However, existing white-box and black-box test selection methods prove ineffective when testing third-party pre-trained DNN models, where internal model details, such as hidden and softmax layers, are typically inaccessible. Consequently, an effective test prioritization approach must operate without requiring access to the internal structure of DNN models.

\begin{boxH}
In summary, testing DNNs presents two primary challenges: (1) executing all possible test inputs incurs high computational costs, and (2) existing selection methods for prioritizing a smaller test set require access to the model's internal structure. 
\end{boxH}

LBT addresses one of these challenges by leveraging the intrinsic relationship between testing and model inference within a program~\cite{walkinshaw2010increasing}. It operates by iteratively training a behavioral model that locally approximates the target system, capturing its decision boundaries. This behavioral model is refined through synthetic counterexamples generated by encoding the behavioral model and utilizing the encoded model with an SMT solver~\cite{sharma2021mlcheck}. It enables the generation and selection of new test cases without requiring knowledge of the internal workings of the System Under Test (SUT).  

However, encoding the behavioral model introduces certain limitations. If the model is overly complex, encoding it becomes computationally expensive. Traditional LBT methods depend on SMT solvers with the encoded model to generate synthetic inputs, further increasing computational costs. Additionally, simplified behavioral models, such as Decision Trees or basic Neural Networks, may fail to accurately capture the target system's behavior or may not be suitable for certain scenarios. Furthermore, due to the encoding process, the behavioral model, an inherent byproduct of the LBT approach, cannot integrate other test generation techniques that function independently of the behavioral model.

\begin{boxH}
To overcome these limitations, we propose an alternative LBT approach that incorporates behavioral model mutation~\cite{ma2018deepmutation} and statistical hypothesis testing~\cite{wald2004sequential}.
\end{boxH}

In this approach, instead of encoding the behavioral model and solving it with an SMT solver, the behavioral model generation utilizes a Jacobian-based heuristic to efficiently explore the input domain of the Model Under Test (MUT). This method identifies directions in which the model's output varies around an initial set of training points. The heuristic prioritizes samples capturing the target DNN's output variations, ensuring that the behavioral model approximates the target's decision boundaries. The process begins with a small initial set of inputs that represent the input domain. 

After training the behavioral model, we generate multiple mutations by applying a set of mutation operators proposed in~\citet{ma2018deepmutation} to assess the effectiveness of the generated test suite. Once the mutated behavioral models are created, we prioritize adversarial inputs independently of the MUT by computing their mutation score on these behavioral model mutants. Specifically, for a given input, we measure how many of the generated mutants of the behavioral model produce an output different from the original behavioral model’s output. Since the behavioral model approximates the decision boundary of the actual MUT, \textit{we hypothesize that adversarial inputs with a higher mutation score on the mutated behavioral models are more likely to be misclassified by the MUT, as these inputs exhibit high sensitivity to model mutations.} Consequently, we selectively execute adversarial inputs with a higher likelihood of being misclassified, thereby reducing the cost of testing the MUT and effectively addressing the second challenge in DNN testing.

To measure the effectiveness of the proposed method, we pose the following research questions, organized around the key aspects of our investigation:

\textbf{- RQ1: Does the proposed method effectively identify a subset of tests with higher fault-detection potential compared to baseline methods?}

\textbf{- RQ2: Does the proposed method generate a more effective test permutation than the baseline methods?}

\textbf{- RQ3: Does the proposed method select test cases that uncover a broader range of faults?}

\textbf{- RQ4: Does the proposed method effectively guide DNN retraining to improve its accuracy?}


\section{Related Works}
\subsection{Test case Prioritization}
Selecting and prioritizing test cases are key aspects of reliability evaluation, helping to lower labeling costs and boost the testing efficiency of DNN systems~\cite{shi2021empirical}. Test prioritization was introduced by \citet{wong1995effect}, which aims to determine the optimal sequence or subset of tests, allowing software testers or developers to maximize their results within a constrained time frame~\cite{feng2020deepgini}. 
Existing test selection approaches for DNN models can be classified as either black-box or white-box, depending on their need for access to the internal workings of the DNN. 

\subsubsection{\textbf{Black-box}}
\citet{feng2020deepgini} introduced DeepGini, a test selection method based on
Gini scores, 
has been demonstrated to be more efficient in detecting mispredictions compared to random and coverage-based selection techniques~\cite{feng2020deepgini,weiss2022simple}. 
\citet{li2019boosting} presented cross-entropy-based sampling and confidence-based stratified sampling for black-box DNN test selection. These metrics are utilized to choose a small subset of test inputs. \citet{arrieta2022multi} conducted metamorphic testing on DNN models using uncertainty metrics. They utilized Non-Dominated Sorting Genetic Algorithm (NSGA)-II along with uncertainty scores to choose metamorphic follow-up inputs to test DNN models. \citet{ma2021test} utilized maximum prediction probability score to prioritize test cases. 

\subsubsection{\textbf{White-Box}}
\citet{pei2017deepxplore} introduced DeepXplore, the first white-box testing framework designed to detect and generate corner-case inputs that could cause varying behaviors across multiple deep neural networks (DNNs). They proposed Neuron Coverage to prioritize test inputs for DNN based on the neuron activation ranges. \citet{ma2018deepgauge} proposed DeepGauge, which introduces multi-granularity testing criteria to evaluate test adequacy from two perspectives: the neuron level and the layer level. 
\citet{wang2021prioritizing} introduced a novel selection metric named PRIMA that employs both image mutation and DNN model mutation to identify data likely to be misclassified by the model, aiming to uncover DNN bugs.
\citet{kim2019guiding} proposed surprise coverage metrics, which measure the relative novelty or unexpectedness of a specific test case compared to the test cases employed during training.  They introduced two test adequacy criteria, LSA and DSA, which we will use as baseline methods for comparison with the proposed prioritization technique. DeepXplore~\cite{pei2017deepxplore}, 
DeepGauge~\cite{ma2018deepgauge}, LSA \& DSA~\cite{kim2019guiding} rely on achieving higher coverage of DNN architectures which serves as a valuable indicator of the test cases' effectiveness~\cite{ma2021test}. 

\subsection{Learning-Based Testing}
\begin{figure}
  \centering
  \includegraphics[width=0.9\columnwidth]{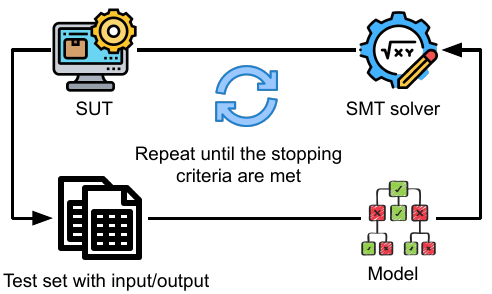}
  \caption{The link between ‘learning’ and ‘testing’ in LBT.}
 \vspace{-5mm}
  \label{fig:basic_lbt}
\end{figure}

The origins of LBT can be traced back to the works of Budd and Angluin~\cite{budd1982two} and Weyuker~\cite{weyuker1983assessing}, who identified two key challenges in program testing that LBT aims to address: 1) choosing test cases that are highly effective in detecting faults, and 2) establishing a stopping criterion to determine when testing has been sufficient to consider the SUT adequately validated. 
They introduced LBT which treats testing as the process of inferring software attributes by observing its responses to inputs~\cite{fraser2012behaviourally,sharma2021mlcheck,papadopoulos2015black}. Figure~\ref{fig:basic_lbt} represents the basic concept of LBT. It builds a behavioral model from test cases, which then guides further test generation~\cite{papadopoulos2015black,walkinshaw2017uncertainty,fraser2012behaviourally}. The model is iteratively refined by identifying counterexamples—inputs that contradict it—helping it converge and directing testing toward unexplored behaviors~\cite{papadopoulos2015black,sharma2021mlcheck}.
Test case generation in LBT often combines machine learning, constraint solving, and formal verification~\cite{sharma2021mlcheck,sharma2020higher,meinke2010learning}. A common approach involves creating a white-box model of a black-box SUT, converting it into logical formulas, and using SMT solvers to identify inputs that violate properties like fairness or monotonicity~\cite{sharma2021mlcheck,sharma2020higher,papadopoulos2015black}. These candidate counterexamples are validated against the actual SUT before being added to the test suite~\cite{fraser2012behaviourally}. The white-box model can be a Decision Tree~\cite{papadopoulos2015black,fraser2015assessing,briand2009using} Abstract Syntax Tree~\cite{walkinshaw2017uncertainty}, overlapping polynomial functions~\cite{meinke2010learning} or even Neural Network~\cite{sharma2020higher}.

 \vspace{-2mm}
\section{Methodology}
\begin{figure*}
  \centering
\fbox{\includegraphics[width=\textwidth]{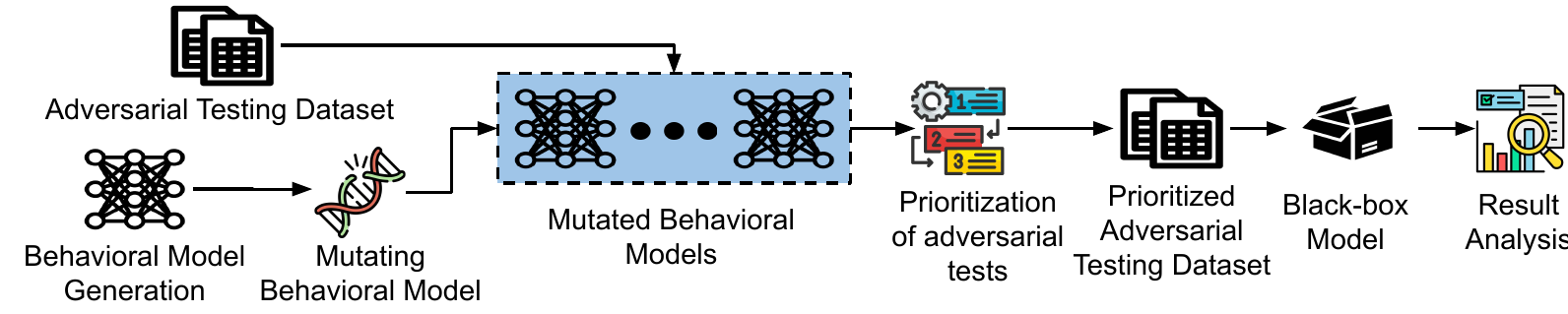}}
  \caption{Overview of the proposed methodology}
   \vspace{-5mm}
\label{fig:process_diagram}
\end{figure*}

We outline our methodology in this section. 
\textbf{Figure~\ref{fig:process_diagram}} provides an overview of the experimental process carried out in this research. \textbf{We initially constructed a behavioral model of the MUT by augmenting input set and progressively learning its decision boundary. Once the behavioral model was generated, we created its mutants and ranked the adversarial inputs according to their mutation score using hypothesis testing.}
\subsection{Behavioral Model Generation}
\textbf{Algorithm~\ref{alg:behavioral_model_training}} defines the behavioral model generation technique. The training process begins by selecting a base model as the behavioral model and labeling an initial dataset \( X \) using the MUT's predictions \( Y \). We define the set of these input-output pairs as \( S \). Additionally, before any training iteration begins, we collect the MUT's outputs for a validation set \( X_{val} \) and store them as \( Y_{val} \). Since the validation set does not require ground truth labels, we perform validation by comparing the predictions of the behavioral model with \( Y_{val} \), measuring the percentage similarity between the MUT and the behavioral model's outputs. Using this similarity score, we quantify how closely the behavioral model replicates the MUT's decision boundary in each iteration. 

We trained the behavioral model in an iterative approach, stopping when the similarity percentage exceeded a predefined threshold or if we observed no improvement for a given number of epochs (patience). In each iteration, we train the behavioral model on the dataset \( S \). After training, we apply Jacobian-based dataset augmentation to generate perturbed versions of the training samples, aiding in refining the behavioral model's decision boundaries. However, instead of adding all augmented samples, we chose only those where the behavioral model and the MUTs disagree and retrained the behavioral model. We added these disagreement samples, along with their corresponding oracle-generated labels, to \( S \), progressively enriching the dataset with challenging examples. We repeated this process iteratively, to retrain the behavioral model on the expanding dataset. Over multiple training cycles, the behavioral model incrementally improves in mimicking the oracle’s behavior, as reflected by an increasing validation similarity percentage.

\begin{algorithm}
\caption{Behavioral Model Generation}
\label{alg:behavioral_model_training}
\begin{spacing}{0.9}
\begin{algorithmic}
\Require Behavioral model $B$, MUT $M$, Initial training set $X$, Validation set $X_{val}$, Similarity threshold $\tau$, Patience $p$
\State  $Y \gets M(X)$, $Y_{val} \gets M(X_{val})$ \Comment{Obtain MUT's predictions for $X$ and $X_{val}$}
\State $S \gets (X,Y)$ \Comment{Initialize training set}
\State $best\_similarity \gets 0$, $no\_improvement \gets 0$

\While{True} 
    \State Train $B$ on $S$
    \State $sim \gets \text{Similarity}(B(X_{val}), Y_{val})$
    \If{$sim > \tau$}
        \State \Return $B$ \Comment{Terminate if similarity threshold is met}
    \EndIf
    \If{$sim > best\_similarity$}
        \State $best\_similarity \gets sim$
        \State $no\_improvement \gets 0$
    \Else
        \State $no\_improvement \gets no\_improvement + 1$
    \EndIf
    \If{$no\_improvement \geq p$}
        \State \Return $B$ \Comment{Terminate if no improvement for $p$ epochs}
    \EndIf
    \State Generate Jacobian-based augmentations for $S$ 
    \State Identify disagreement samples $S_d$ where $M$ and $B$ disagree
    \State $S \gets S \cup S_d$ \Comment{Expand training set}
\EndWhile
\end{algorithmic}
\end{spacing}
\end{algorithm}

\subsection{Prioritization Technique}
Next, we start the prioritization process. Algorithm~\ref{alg:mutation-selection} shows the prioritization process in detail. 
\subsubsection{\textbf{Setting Mutation Rate \& Operator}}
After generating the behavioral model, we produced mutant versions of each behavioral model using mutation operators introduced in~\citet{ma2018deepmutation}. 
Since DNNs differ significantly from traditional software systems, conventional mutation operators can not be directly applied. Ma et al.~\cite{ma2018deepmutation} introduced a set of mutation operators for DNN-based systems at different levels, including source-level (e.g., training data and programs) and model-level (e.g., the DNN model itself). In our approach, we excluded source-level mutations, as retraining behavioral models from scratch would be time-consuming, and instead focused on model-level operators that directly modified the behavioral model. Specifically, we adopted three of the eight operators proposed by~\citet{tian2018deeptest}.

We created multiple mutated variations of the behavioral model. Then we ranked adversarial inputs based on the number of these mutants that generate outputs differing from the original behavioral model's response to the same input. As the target was mutating the behavioral model, we focused on the model level operators that modified the behavioral model to obtain mutated versions of the behavioral model as the behavioral model represents the decision boundary of the MUT. \textit{We hypothesized that if the adversarial input has output on most of the mutated models that is different from the output of the behavioral model, the adversarial input has high possibility of getting misclassified by the MUT}. 

However, we ensured that a mutated model may not behave much differently than the original behavioral model during mutation. Otherwise, the mutated model may generate a different input than the original behavioral model even for a benign input. Using various mutation operators, we first determined an appropriate mutation rate at which the behavioral model undergoes mutation. The mutation rate is chosen based on the recommendations by~\citet{ma2018deepmutation} to avoid this issue. Additionally, this selection ensured an effective mutation score distance between adversarial inputs prone to misclassification and those likely to be correctly classified. 
We primarily focused on four model-level mutation operators to mutate the behavioral model: Gaussian Fuzzing (GF), Weight Shuffling (WS), Neuron Activation Inverse (NAI), Neuron Switch (NS), as discussed by~\citet{ma2018deepmutation}. We did not use the other 4 model level operators because they would alter the structure of the behavioral model. However, for this experiment, we chose to preserve the architecture of the behavioral model because changing the structure may result in generation of invalid mutants of the behavioral model. Among the four selected model-level operators, WS was ineffective in generating meaningful mutants, as they exhibited behavior nearly identical to the behavioral model, making input differentiation impossible. Therefore, we used the remaining three operators GF, NAI, and NS to generate mutants.
\subsubsection{\textbf{Setting Mutation Score Threshold}}
Before the final prioritization technique starts in \textbf{step 3}, we created 100 mutants for a behavioral model, in \textbf{step 1} of the \textbf{algorithm~\ref{alg:mutation-selection}}. We executed the validation set $X_{val}$ used to calculate the similarity between the behavioral model and the MUT on these mutants and calculated the average percentage of mutation score \(\zeta_h\) for a normal input, which is the mutation score threshold that we expected for normal (non-adversarial) samples. An adversarial input should have a mutation score more than this threshold to be selected.
\subsubsection{\textbf{Prioritization of Adversarial Inputs}}

Our proposed method prioritizes inputs that yield the most diverse outputs across mutated models compared to the behavioral model. While executing every input on all mutated models ensures thorough analysis, it is computationally expensive. Instead, we focus on adversarial inputs, ranking them based on mutation scores while avoiding exhaustive execution. 
For instance, if an input produces the same output across 90 out of 100 mutated models, it is likely insensitive to mutations, making further execution unnecessary. Conversely, if an input behaves differently in those 90 models, it is highly sensitive and should be prioritized. The challenge is determining when to stop execution and how to order inputs to maximize misclassification potential. Our approach selectively picks and ranks inputs based on partial mutation scores, ensuring the most impactful inputs are tested first.


To tackle these challenges, we used Sequential Probability Ratio Test (SPRT)~\cite{wald2004sequential} on each adversarial input from which our proposed method will prioritize test cases. We used it because the sequential nature makes SPRT particularly valuable when data comes in over time, decisions need to be made quickly, or when the cost of collecting data is high. SPRT is a statistical method used for testing two hypotheses—typically a null hypothesis $H_0$ and an alternative hypothesis $H_1$ sequentially. Instead of fixing a sample size in advance, SPRT evaluates data as collected and stops as soon as there is enough evidence to decide for any specific data point.

In the \textbf{Step 3} of \textbf{algorithm~\ref{alg:mutation-selection}}, adversarial inputs were executed on a new mutant model to calculate the SPRT value. Based on the output, whether to select or keep them unselected was decided. The adversarial input that was selected early was prioritized to be executed on the MUT. The method iteratively selected a mutated behavioral model unless the list of adversarial inputs is empty or all the mutated models have been picked.
We fixed the total mutants because otherwise, the loop may run infinitely if some input may fall between the threshold $A$ and $B$. To avoid this case, we proposed to use a fixed number of mutated behavioral model $n_{max}$. The iterative loop created mutated behavioral models and calculated likelihood ratio for each input as long as $n$<$n_{max}$.
In this way, we avoided the possibility of the loop running infinitely. The inputs that could not be selected from the adversarial list after $n_{max}$ iteration were discarded. 

\textbf{Step 2} in \textbf{algorithm~\ref{alg:mutation-selection}} shows how the number $n_{max}$ is selected and how total $n_{max}$ amount of mutants are created.
To determine the maximum number of mutated models, \( n_{\max} \), we randomly selected a small subset of adversarial inputs. A loop is then executed where mutated models are generated from these inputs, and the SPRT algorithm is applied to evaluate them. This process continues until 90\% of the inputs have either been selected or discarded. If even after executing 100 initially created mutants, the 90\% criteria is not filled up, new mutants are created and added to the mutants list, and thus the process continues. Once the loop completes, the total number of mutated models generated serves as \( n_{\max} \), and any remaining adversarial inputs from the initial set are discarded.

For the rest of the adversarial inputs, a second loop runs in \textbf{Step 2}, but this time using the fixed \( n_{\max} \). This approach allows us to determine an appropriate \( n_{\max} \) using a smaller subset of inputs before applying it consistently to the rest of the large dataset.

In this experiment, the parameters \( A \), \( B \), and \( \delta \) which are required by the SPRT algorithm, set at 5\% to minimize the chances of Type I and Type II errors. This strict threshold makes adversarial input selection more competitive, as only those with a high mutation score will be chosen.


\begin{algorithm}
\caption{Test Prioritization Technique}
\label{alg:mutation-selection}
\begin{spacing}{0.75}
\begin{algorithmic}
\State \textbf{Input:} Set of adversarial inputs $X_{adv}$, validation dataset $X_{val}$, mutation operator $\mathcal{M}$, $\alpha$, $\beta$, $\delta$
\State \textbf{Output:} Selected adversarial inputs for testing, $X_{sel}$

\State \textbf{Step 1: Compute mutation score threshold} \;
\State Generate 100 mutated models using $\mathcal{M}$\;
\State Compute mutation scores for benign inputs $x \in X_{val}$;
\State Calculate average mutation score threshold $\zeta_h$\;

\State \textbf{Step 2: Creating mutants and setting $n_{max}$}
\State Randomly select a small set of adversarial inputs $X_0 \subset X_{adv}$
\State Initialize $n_{\max} \gets 100$,  discarded count $d \gets 0$, $n \gets 1$, $X_{sel} \gets \{\}$

\While{$d < 0.9 \times |X_0|$}
    \If{$n \leq n_{max}$}
        \State Get model $M'_n$ from $\{M'_1, M'_2, ..., M'_{n_{max}}\}$
    \Else
        \State Generate a mutated model $M'_n$ using operator $\mathcal{M}$
        \State $n_{max} = n_{max} +1 $
        \State Append the new model $M'_n$ in the mutated models list
        
    \EndIf
    \State $n = n+1$
    \For{each adversarial input $x \in X_0$}
        \If{output of $M'_n$ differs from the original model}
            \State Increment mutation score $z$ for $x$
        \EndIf
        \State $S = \text{SPRT}(\zeta, \textit{n}, \textit{z}, \delta)$
\If{$S \geq  \frac{1-\beta}{\alpha}$}
    \State Discard $x$ from $X_0$
    \State $d = d+1$
\ElsIf{$S \leq \frac{\beta}{1-\alpha}$}
    \State Discard $x$ from $X_0$ and append $x$ to $X_{sel}$ 
    \State $d = d+1$
\EndIf

\EndFor
\EndWhile
\State \textbf{Step 3: Select adversarial inputs with fixed $n_{\max}$}
\For{each mutated model $M'_n$ in $\{M'_1, M'_2, ..., M'_{n_{\max}}\}$}
    \For{each adversarial input $x \in X_0 \setminus X_{adv}$}
        \If{output of $M'_n$ differs from the original model}
            \State Increment mutation score $z$ for $x$
        \EndIf
        \State $S = \text{SPRT}(\zeta, \textit{n}, \textit{z}, \delta)$
        \If{$S \geq  \frac{1-\beta}{\alpha}$}
            \State Discard $x$ from $X_0 \setminus X_{adv}$
        \ElsIf{$S \leq \frac{\beta}{1-\alpha}$}
            \State Discard $x$ from $X_0 \setminus X_{adv}$ and append in $X_{sel}$
        \EndIf
    \EndFor
\EndFor
\State \textbf{Return:} Selected adversarial inputs $X_{sel}$

\end{algorithmic}
\end{spacing}
\end{algorithm}

Once prioritization is complete, the number of selected adversarial inputs is determined. An equivalent number of test cases are then extracted from the ranked lists of the baseline methods. We use these cases to compare approaches using four metrics to evaluate our method's effectiveness.

\section{Baseline Methods}
We assessed our proposed LBT-based selection method by comparing it with eight baseline approaches, which include various types of test case prioritization and detection techniques:  one distributional detection method Kernel Density Estimation (KDE)~\cite{feinman2017detecting},
one surprise-adequacy based method Distance-based Surprise Adequacy (DSA)~\cite{kim2019guiding}, one Normalization based detection method Dropout Randomization (DR)~\cite{feinman2017detecting}, two well-known coverage-based approaches Neuron Activation Coverage (NAC) and Neuron Boundary Coverage (NBC)~\cite{pei2017deepxplore}, three well-known confidence dispersion metrics DeepGini~\cite{feng2020deepgini}, PE~\cite{feng2020deepgini} \& MaxP~\cite{ma2021test}, and random selection. 
Among these baselines, KDE, DSA, DR, NAC, and NBC are classified as white-box methods since they require access to the neuron activation values from the hidden layers of DNNs. On the other hand, DeepGini, MaxP, and PE are considered black-box methods because they do not require such access, although they do rely on the softmax layer to obtain probability distributions.
Our selection of baselines is guided by prior research on adversarial input detection and test case prioritization by~\citet{shi2021empirical,feng2020deepgini,aghababaeyan2024deepgd}, and~\citet{pei2017deepxplore}.
For DR, the choice of \( L \),  the number of times inputs are applied through a randomized network,  is not highly sensitive, but results are stable for \( L > 20 \). Based on the experiments in previous study~\cite{carlini2017adversarial}, we set \( L = 30 \) in this experiment.
Following prior experiments~\cite{kim2019guiding,shi2021empirical}, we computed DSA and KDE using the final hidden layer of the MUT. For NAC and NBC, we adopt the parameter values from~\citet{feng2020deepgini}, as the authors deemed them effective. We ranked the adversarial inputs based on the outputs of baseline methods and sorted in descending or ascending order, depending on the requirement.

\section{Experimental Setup}



\subsection{Model Architectures}
\label{model_architectures}
To demonstrate the effectiveness of the proposed method for selecting adversarial inputs, the selection of the type of DNN as MUT is crucial. 
We employed two distinct deep learning (DL) model architectures, namely \textbf{ResNet-20}~\cite{jinliang2019cifar10} and \textbf{VGG-16}~\cite{simonyan2014very}, as these architectures are commonly used as MUTs in existing DNN test case prioritization baseline methods~\cite{he2016deep,shi2021empirical,aghababaeyan2024deepgd,gao2022adaptive}. We trained, tested, and evaluated these models with different combinations to assess the method's performance on a large benchmark dataset. For each of the model architectures, we trained two different DNN models using the two selected datasets. 
We trained a total of four DNN models, which serve as the MUTs in this experiment. 
\textbf{Table~\ref{tab:muts}} presents the MUTs and their respective accuracy in each benign validation set and corresponding adversarial set.

To demonstrate the model mutation techniques, we used \textbf{LeNet-5}~\cite{sun2018convolutional} architecture as the base for the behavioral model. We used it with fewer layers and complexity than the MUTs as a similar model was used in~\citet{ma2018deepmutation}.

\subsection{Datasets}
\label{dataset}
We selected two widely recognized publicly available datasets, MNIST~\cite{deng2012mnist} and Fashion-MNIST~\cite{xiao2017fashion}, as evaluation subject datasets. These datasets have been extensively utilized in numerous baseline method experiments~\cite{feng2020deepgini,aghababaeyan2024deepgd,gao2022adaptive,shi2021empirical}.

\subsection{Adversarial Example Generation}
We selected two cutting-edge adversarial attack tools: \textbf{FGSM} (Fast Gradient Sign Method)~\cite{goodfellow2014explaining} and  
and
\textbf{JSMA} (Jacobian Saliency Map Attack)~\cite{papernot2016limitations} 
to create adversarial examples. We utilized the existing Python library, Adversarial Robustness Toolbox (ART)~\cite{art2018} to execute these attacks, configuring each attack using Manual settings.
We selected these two evasive methods because they are particularly effective for real-time, post-deployment attacks aimed at misleading a model's predictions without modifying the model itself. Evasive techniques primarily require access only to the model's predictions, which aligns well with our experiment, as we consider the MUT to be a black-box. These techniques take the MUT as a parameter and can generate adversarial datasets based on the provided benign training or testing data. 

As discussed in the sections~\ref{model_architectures} and~\ref{dataset}, there are four MUTs based on two model architectures trained on two different datasets. For each MUT, we generated two adversarial testing datasets for two corresponding adversarial attacks. Therefore, we generated a total of eight adversarial testing sets based on all the possible combinations of MUT, dataset, and adversarial generation techniques. \textbf{Table~\ref{tab:muts}} displays all the MUTs, their overall accuracy, and faults on the corresponding benign and adversarial test sets generated using selected adversarial attacks.

We aimed to maintain the accuracy on adversarial test cases within a specific range: \textit{it should not fall below 40\%, and it should be at least 30\% lower than the accuracy achieved by the MUT on the corresponding benign test dataset}. For instance, \textbf{Table~\ref{tab:muts}} shows that the ResNet-20 architecture model trained on the MNIST dataset achieves an accuracy of 96.56\% on the benign test set. However, when the same model is evaluated on an adversarial test set generated using the FGSM attack, its accuracy drops to 58.17\%. This range was essential for accurate evaluation. If the accuracy of adversarial test cases is too low, it becomes difficult to assess how our proposed method effectively selects adversarial inputs from a mixed set of benign and adversarial samples. Adversarial test cases generated by these functions, which are near the classifier's decision boundary, might lack meaningful characteristics, making it difficult for even human annotators to accurately determine its category~\cite{zhan2022comparative}. Even images produced by generative adversarial networks, such as GAAL~\cite{zhu2017generative}, encounter this problem. 
Contrarily, if the accuracy of adversarial inputs exceeds or matches that of benign inputs, the model might prioritize benign inputs over true adversarial ones due to its higher confidence in the adversarial examples. 

Therefore, for the total eight combinations of MUT, dataset, and adversarial generation technique, we manually tuned the parameters across multiple iterations and checked the overall accuracy of each model for the generated set of adversarial inputs. This process ensured that the adversarial examples were sufficiently challenging while providing a meaningful context for evaluating the performance of our proposed approach.

\begin{table}
\centering
  \raisebox{0.05\height}{ 
        \scalebox{0.9}[0.9]{ %
\begin{tabular}{|c|c|c|c|c|}
\hline
\multirow{2}{*}{\textbf{Dataset}} & \multirow{2}{*}{\textbf{Model}} & \multicolumn{3}{c|}{\textbf{Accuracy}}\\ 
\cline{3-5}
 &  & \textbf{Benign} & \textbf{FGSM} & \textbf{JSMA} \\
\hline
MNIST & ResNet-20  
& 96.56\%  & 58.17\% & 59.24\%\\
      & VGG-16   & 99.35\%  & 60.98\% & 59.57\%\\
\hline
FashionMNIST & ResNet-20 & 89.58\% & 69.42\%  & 71.89\%\\
      & VGG-16 & 92.67\%  & 75.28\%  & 73.38\%\\
\hline
\end{tabular}
}
}
\caption{Accuracy and of MUTs on Benign and Adversarial Test Set Generated Using ART Library}
\vspace{-6pt}
\label{tab:muts}
\end{table}

\subsection{System Configuration}
We implemented the proposed test generation and baseline methods in Python (3.10.10) based on Keras~\cite{chollet2021deep} (2.12.0) with TensorFlow~\cite{tensorflow2015-whitepaper} (2.12.0) as backend. For calculating DSA and KDE for a test input, we used dnn-tip~\cite{10.1145/3533767.3534375} (0.1.1) python library.
We ran experiments on a Rocky Linux 8.10 server with a 28-core 2.20GHz Xeon CPU and two NVIDIA L40 GPUs (64GB).

\section{Results}
\subsection{RQ1:  Fault detection}
To answer this research question, we utilized Fault Detection Rate (FDR)[equation~\ref{eq:fdr}] as the metric to evaluate the effectiveness of the proposed prioritization technique. The purpose of the FDR metric is to quantify the proportion of faults that a test set can detect relative to the maximum possible faults it could potentially uncover, considering either the size of the test set or the total number of faults in the dataset. This metric ranges from 0 to 1.
The FDR has been utilized as an evaluation metric to assess the effectiveness of several test prioritization techniques~\cite{gao2022adaptive,aghababaeyan2024deepgd,shi2021empirical}. 

FDR can be expressed as:
\begin{equation}
\small {FDR}(S) = \frac{|F_S|}{\min(|S|, |F|)}
\label{eq:fdr}
\end{equation}
\textit{Here, S  is the selected test input subset, $F_S$ \text represents the number of possible faults revealed by  S, \ |S| \text{ is the size of the test input set, and } |F| is the total number of faults detected by the entire dataset.} 

We utilized the ART library to generate adversarial inputs from the benign test set. We must determine the number of faults the MUT can produce to calculate FDR. 
Adversarial inputs are generated based on the benign validation set and we already have the oracle for each test input, which is essential for computing the FDR.

\textbf{Table~\ref{tab:fdr}} presents the average FDR of the selected subsets for the proposed LBT method alongside baseline methods across various combinations of datasets, model architectures, and adversarial attack techniques. The BM-MO column represents the mutation operator applied to modify the behavioral model. This convention is consistently used in the other evaluation tables.

The results clearly demonstrate that our proposed LBT method offers a more effective prioritization strategy than the baseline methods, as the FDR of the subset prioritized by our method with LBT consistently surpassed that of the other baseline methods more frequently.
This observation indicates that our proposed method consistently enhances the identification of relevant adversarial inputs, highlighting its robustness in the prioritization process for adversarial test inputs. 
Among the baseline methods, KDE and DSA exhibited the poorest performance, as the average FDR for the subsets selected by these methods across all combinations was consistently lower even than that achieved through random selection. Following our proposed method, the next most effective methods were DR and confidence-based metrics DeepGini, PE,  and MaxP. 

\begin{table*}
    \centering
         \raisebox{0.05\height}{ 
        \scalebox{0.85}[0.85]{ 
    \begin{tabular}{|l|l|l|l|l|l|l|l|l|l|l|l|l|l|}
    \hline
\textbf{Attack} & 
\textbf{BM-MO} & 
\textbf{MUT} & 
\textbf{LBT} & 
\textbf{KDE} & 
\textbf{DR} & 
\textbf{DSA} &
\textbf{NAC} & 
\textbf{NBC} & 
\textbf{DeepG.} & 
\textbf{Maxp} & 
\textbf{PE} & 
\textbf{Rand.} \\  
\hline
\multicolumn{13}{|c|}{MNIST}\\
\hline
        FGSM & GF & ResNet-20 & \textbf{0.9474} & 0.1818 & 0.8649 & 0.3909 & 0.8649 & 0.6706 & 0.9404 & 0.9404 & 0.6955 & 0.8710 \\
       
        & & VGG-16 & \textbf{0.9994} & 0.5189 & 0.8756 & 0.7461 & 0.9418 & 0.9418 & 0.9991 & 0.9751 & 0.9991 & 0.9204 \\
       
        & NAI & ResNet-20 & \textbf{0.9627} & 0.1554 & 0.9269 & 0.3936 & 0.9269 & 0.593 & 0.9467 & 0.9477 & 0.6592 & 0.9161 \\
        
        & & VGG-16 & \textbf{0.9383} & 0.4811 & 0.8812 & 0.7368 & 0.9354 & 0.9354 & 0.8983 & 0.9108 & 0.9183 & 0.8761 \\
       
         & NS & ResNet-20 & 0.8917 & 0.1772 & 0.9647 & 0.3922 & \textbf{0.9647} & 0.6628 & 0.9419 & 0.9419 & 0.6944 & 0.8122 \\

        &  & VGG-16 & \textbf{0.9346} & 0.5006 & 0.8793 & 0.7408 & 0.9275 & 0.8355 & 0.9187 & 0.9275 & 0.9187 & 0.9262 \\ \hline
        
         JSMA & GF & ResNet-20 & 0.9155 & 0.3811 & \textbf{0.9583} & 0.7847 & 0.8583 & 0.7593 & 0.9218 & 0.9520 & 0.8645 & 0.8165 \\ 
         & & VGG-16 & 0.8301 & 0.2542 & 0.8569 & 0.5484 & 0.8107 & 0.8107 & \textbf{0.9371} & 0.8352 & 0.9314 & 0.9283 \\ 
        & NAI & ResNet-20 & 0.8974 & 0.3944 & 0.9525 & 0.8122 & 0.9525 & 0.9540 & 0.9676 & 0.9563 & 0.963 & \textbf{0.9691} \\
         & & VGG-16 & \textbf{0.9618} & 0.2218 & 0.8535 & 0.5503 & 0.8035 & 0.8035 & 0.9268 & 0.9330 & 0.9242 & 0.8706 \\
        & NS & ResNet-20 & \textbf{0.9612} & 0.3795 & 0.8620 & 0.7704 & 0.8620 & 0.9600 & 0.9198 & 0.9506 & 0.9194 & 0.8755 \\ 
        & & VGG-16 & 0.7725 & 0.2705 & 0.8578 & 0.5473 & 0.8060 & 0.8060 & 0.9379 & \textbf{0.9393} & 0.9365 & 0.8515 \\ \hline

\multicolumn{13}{|c|}{FashionMNIST}\\
\hline
        
        FGSM & GF & ResNet-20 & \textbf{0.9369} & 0.1662 & 0.8661 & 0.6183 & 0.7461 & 0.8461 & 0.8706 & 0.9178 & 0.8646 & 0.8743 \\ 
     
        & & VGG-16 & \textbf{0.9493} & 0.033 & 0.6591 & 0.7795 & 0.9113 & 0.9113 & 0.8086 & 0.8378 & 0.7959 & 0.7681 \\

        & NAI & ResNet-20 & 0.9261 & 0.1635 & \textbf{0.9471} & 0.6170 & 0.9471 & 0.9471 & 0.8689 & 0.9368 & 0.8706 & 0.9227 \\ 
       
        &  & VGG-16 & \textbf{0.9315} & 0.0641 & 0.7670 & 0.6746 & 0.9045 & 0.9045 & 0.8624 & 0.8750 & 0.8549 & 0.9083 \\

        & NS & ResNet-20 & 0.9088 & 0.1772 & \textbf{1.0033} & 0.6272 & 0.9533 & 0.9163 & 0.9091 & 0.9948 & 0.9601 & 1.0579 \\ 
        
        & & VGG-16 & \textbf{0.8949} & 0.0594 & 0.7577 & 0.6901 & 0.8051 & 0.9051 & 0.8567 & 0.8696 & 0.8471 & 0.7788 \\ \hline

           JSMA & GF & ResNet-20 & 0.8506 & 0.1147 & 0.9157 & 0.5588 & 0.9157 & 0.864 & 0.9363 & 0.9433 & \textbf{0.9575} & 0.9271 \\ 
           
          & & VGG-16 & 0.8826 & 0.1066 & 0.8143 & 0.5470 & 0.9209 & 0.8801 & 0.9276 & \textbf{0.9367} & 0.9242 & 0.9317 \\ 
         & NAI & ResNet-20 & \textbf{0.8907} & 0.1209 & 0.8594 & 0.5386 & 0.7194 & 0.8687 & 0.8343 & 0.8454 & 0.8608 & 0.8720 \\
         &  & VGG-16 & 0.8535 & 0.1493 & 0.8352 & 0.4766 & 0.9224 & 0.8819 & 0.9131 & \textbf{0.9454} & 0.9143 & 0.8971 \\ 
        & NS & ResNet-20 & 0.8684 & 0.1449 & 0.9159 & 0.5141 & 0.9159 & 0.8666 & 0.9377 & 0.9422 & \textbf{0.9637} & 0.9088 \\ 
         & & VGG-16 & \textbf{0.9481} & 0.1586 & 0.8794 & 0.4867 & 0.8699 & 0.9293 & 0.9128 & 0.9240 & 0.9109 & 0.8866 \\ \hline
    \end{tabular}
    }
    }
    \caption{FDR of the test set prioritized by LBT and baseline approaches from adversarial dataset generated from adversarial attack FGSM and JSMA on MUTs trained on MNIST and FashionMNIST}
\label{tab:fdr}
\end{table*}

\begin{dashedbox}
Overall, the test set prioritized by our proposed LBT method demonstrates a significantly higher fault detection capability compared to the subsets selected by the baseline methods.
\end{dashedbox}

\subsection{RQ2: Test permutation}
We address RQ2 by calculating the Average Percentage of Fault Detection (APFD) metric~\cite{yoo2012regression}. Higher APFD values indicate a faster detection rate of misclassified tests. APFD is determined by plotting the percentage of detected misclassified tests against the number of prioritized tests and computing the area under the resulting curve.


The APFD value for this permutation is calculated as:
\begin{equation}
\small {APFD} = 1 - \frac{\sum_{i=1}^{k} o_i}{kn} + \frac{1}{2n}
\end{equation}
\textit{Here, for a permutation of \( n \) tests containing \( k \) misclassified tests, let \( o_i \) represent the position of the first test that identifies the \( i^{th} \) misclassified test. }

To ensure consistency across different scales, we normalized the APFD value from its theoretical minimum (\( \text{min} \)) and maximum (\( \text{max} \)) to a range of \([0, 1]\). With this normalization, a prioritization method is more effective if its APFD value approaches 1 and less effective if it nears 0.

\textbf{Table~\ref{tab:apfd}} presents the average APFD of the selected subsets for our proposed LBT method alongside baseline methods across different combinations of datasets, model architectures, and adversarial attack techniques. In addition to achieving a high FDR score, our proposed method demonstrates superior permutation quality compared to baseline approaches. While white-box methods such as DSA, NAC, and NBC also achieve the highest APFD values in different experiments, our proposed method consistently outperforms them in this metric despite having no access to the internal structure of the MUT.

\begin{table*}
    \centering
     \raisebox{0.05\height}{ 
        \scalebox{0.90}[0.90]{ %
    \begin{tabular}{|l|l|l|l|l|l|l|l|l|l|l|l|l|l|}
    \hline
        \textbf{Attack} & 
\textbf{BM-MO} & 
\textbf{MUT} & 
\textbf{LBT} & 
\textbf{KDE} & 
\textbf{DR} & 
\textbf{DSA} &
\textbf{NAC} & 
\textbf{NBC} & 
\textbf{DeepG.} & 
\textbf{Maxp} & 
\textbf{PE} & 
\textbf{Rand.} \\ \hline
        
\multicolumn{13}{|c|}{MNIST}\\
\hline
         FGSM & GF & ResNet-20 & \textbf{0.5031} & 0.423 & 0.5008 & 0.4871 & 0.4923 & 0.4610 & 0.5011 & 0.5012 & 0.4718 & 0.5000 \\ 
         & & VGG-16 & \textbf{0.5015} & 0.4718 & 0.5000 & 0.4995 & 0.4936 & 0.4936 & 0.4999 & 0.5014 & 0.4999 & 0.4995 \\ 
         & NAI & ResNet-20 & 0.5007 & 0.3935 & 0.5011 & 0.4506 & \textbf{0.5765} & 0.5354 & 0.5039 & 0.5036 & 0.4623 & 0.4994 \\
          & & VGG-16 & \textbf{0.5053} & 0.4732 & 0.4984 & 0.5012 & 0.4957 & 0.4957 & 0.4998 & 0.5003 & 0.4998 & 0.4996 \\ 
         & NS & ResNet-20 & 0.4954 & 0.424 & 0.5010 & 0.4836 & 0.4902 & 0.4945 & 0.5037 & \textbf{0.5039} & 0.4689 & 0.5013 \\ 
         & & VGG-16 & \textbf{0.5051} & 0.4703 & 0.4997 & 0.5043 & 0.4884 & 0.4784 & 0.4998 & 0.5014 & 0.4998 & 0.5018 \\ \hline
         JSMA & GF & ResNet-20 & 0.4976 & 0.5208 & 0.4964 & \textbf{0.5213} & 0.5081 & 0.5085 & 0.4979 & 0.5022 & 0.4991 & 0.4977 \\ 
          & & VGG-16 & \textbf{0.4989} & 0.4392 & 0.4980 & 0.4033 & 0.4641 & 0.4641 & 0.4949 & 0.4974 & 0.4952 & 0.4025 \\ 
         & NAI & ResNet-20 & 0.4962 & 0.5173 & 0.4963 & 0.5169 & 0.5881 & \textbf{0.5887} & 0.4990 & 0.5017 & 0.4997 & 0.5000 \\ 
         & & VGG-16 & \textbf{0.4996} & 0.4492 & 0.4967 & 0.4981 & 0.4632 & 0.4352 & 0.4968 & 0.4978 & 0.4956 & 0.4017 \\ 
         & NS & ResNet-20 & 0.5001 & 0.5149 & 0.4964 & \textbf{0.5207} & 0.5162 & 0.5154 & 0.5000 & 0.5021 & 0.4978 & 0.5015 \\ 
         & & VGG-16 & \textbf{0.4967} & 0.4399 & 0.4985 & 0.4537 & 0.4631 & 0.4631 & 0.4966 & 0.4956 & 0.4946 & 0.4952 \\ \hline

\multicolumn{13}{|c|}{FashionMNIST}\\
\hline

         FGSM & GF & ResNet-20 & 0.4990 & 0.4461 & 0.5004 & \textbf{0.5191} & 0.4858 & 0.4858 & 0.5010 & 0.5006 & 0.4406 & 0.5019 \\ 
         &  & VGG-16 & \textbf{0.4967} & 0.2954 & 0.4590 & 0.4969 & 0.4698 & 0.4698 & 0.4963 & 0.4913 & 0.4944 & 0.4028 \\ 
         & NAI & ResNet-20 & 0.4992 & 0.4609 & 0.4999 & 0.5177 & \textbf{0.5859} & 0.5459 & 0.5018 & 0.5010 & 0.4461 & 0.5017 \\ 
          &  & VGG-16 & 0.4958 & 0.3226 & 0.4561 & 0.5402 & 0.4561 & \textbf{0.5571} & 0.4812 & 0.4882 & 0.4786 & 0.4998 \\ 
         & NS & ResNet-20 & \textbf{0.4975} & 0.4713 & 0.4913 & 0.4280 & 0.4050 & 0.4050 & 0.4966 & 0.4813 & 0.4603 & 0.4689 \\ 
         & & VGG-16 & \textbf{0.5028} & 0.3178 & 0.4540 & 0.5013 & 0.4450 & 0.4350 & 0.4813 & 0.4893 & 0.4795 & 0.5019 \\ \hline
         JSMA & GF & ResNet-20 & 0.4999 & 0.5261 & 0.5024 & \textbf{0.5267} & 0.5113 & 0.4940 & 0.4937 & 0.4974 & 0.4958 & 0.4987 \\ 
          &  & VGG-16 & \textbf{0.5018} & 0.3864 & 0.4971 & 0.4382 & 0.4258 & 0.4131 & 0.5010 & 0.5007 & 0.4958 & 0.5007 \\ 
         & NAI & ResNet-20 & \textbf{0.5019} & 0.4926 & 0.5000 & 0.5013 & 0.4254 & 0.4093 & 0.4970 & 0.4977 & 0.4962 & 0.4991 \\ 
          &  & VGG-16 & 0.4957 & 0.4112 & 0.4917 & 0.5509 & \textbf{0.5680} & 0.5574 & 0.5047 & 0.4986 & 0.5021 & 0.5037 \\ 
          & NS & ResNet-20 & \textbf{0.5054} & 0.4435 & 0.5016 & 0.5019 & 0.4373 & 0.4224 & 0.4976 & 0.5007 & 0.4973 & 0.4991 \\ 
          & & VGG-16 & 0.4937 & 0.4209 & 0.4925 & 0.5577 & \textbf{0.5782} & 0.5685 & 0.5027 & 0.4990 & 0.502 & 0.5019 \\ \hline
    \end{tabular}
    }
    }
        \caption{APFD of the test set prioritized by LBT and baseline approaches from adversarial dataset generated from adversarial attack FGSM and JSMA on MUTs trained on MNIST and FashionMNIST}
        \label{tab:apfd}
\end{table*}

\begin{dashedbox}
Like FDR, the test set prioritized by our proposed LBT method exhibits a significantly higher APFD value, highlighting its ability to arrange the selected inputs so that misclassified cases on the MUT appear earlier in the order.
\end{dashedbox}

\subsection{RQ3: Fault diversity}
Chan et al.~\cite{chen2005adaptive} observed that failure-inducing inputs often cluster closely together where similar faults may stem from the same underlying defect. 
Based on this observation, we aimed for test selection methods that detect more faults and uncover a wider variety of faults efficiently to comprehensively evaluate the MUTs. To address this issue, we adopt the concept of fault type introduced in ~\citet{gao2022adaptive}. Under this metric, for a misclassified test case \(x\), its fault type is defined as:
\[
\small{Fault\_Type}(x) = (\text{Label}^*(x) \rightarrow \small{Label}(x))
\]
where \(\text{Label}^*(x)\) is the ground-truth label, and \(\text{Label}(x)\) is the prediction from the DNN model. For instance, if a handwritten digit with a true label of "7" is misclassified as "1," the fault type is:
\[
\text{Fault\_Type}(x) = (7\!\rightarrow\!1)
\]
Since there are 10 categories in the candidate set, the total number of possible fault types is: 10 $\times 9 = 90$.

To measure the effectiveness of test selection methods in capturing diverse faults, we compute the \textbf{cumulative sum of fault types} identified by each method and compare it to the theoretical maximum curve. We then calculate the \textbf{RAUC (Ratio of Area Under the Curve)}, which quantifies the proportion of the area under the selection method’s curve relative to the theoretical curve. This RAUC score presented in \textbf{Table~\ref{tab:diversity}} helps us assess how well a test selection method covers diverse fault types.

Our proposed method's prioritized inputs exhibit lower diversity compared to the KDE and DSA methods. However, it ranks third in achieving the highest diversity in different dataset and attack combinations, following KDE and DSA. The superior performance of KDE and DSA is attributed to their selection strategy, which primarily relies on ``surprise adequacy'', measuring how different an input is from the training data. This approach ensures high diversity in the selected dataset. Additionally, these methods have full access to the internal structure of the MUT. Despite lacking such access, our proposed method remains effective, outperforming other baselines except for KDE and DSA.

\begin{table*}
    \centering
     \raisebox{0.05\height}{ 
        \scalebox{0.90}[0.90]{ %
    \begin{tabular}{|l|l|l|l|l|l|l|l|l|l|l|l|l|l|}
    \hline
        \textbf{Attack} & 
\textbf{BM-MO} & 
\textbf{MUT} & 
\textbf{LBT} & 
\textbf{KDE} & 
\textbf{DR} & 
\textbf{DSA} &
\textbf{NAC} & 
\textbf{NBC} & 
\textbf{DeepG.} & 
\textbf{Maxp} & 
\textbf{PE} & 
\textbf{Rand.} \\ \hline
        \multicolumn{13}{|c|}{MNIST}\\
\hline
        FGSM & GF & ResNet-20 & 0.3865 & \textbf{0.9500} & 0.3728 & 0.9193 & 0.3728 & 0.1378 & 0.4395 & 0.4377 & 0.4152 & 0.3698 \\ 
        & & VGG-16 & \textbf{0.5722} & 0.5270 & 0.5244 & 0.5423 & 0.5502 & 0.5602 & 0.4764 & 0.4797 & 0.5574 & 0.3864 \\ 
         & NAI & ResNet-20 & 0.3744 & \textbf{0.8597} & 0.3525 & 0.8015 & 0.3525 & 0.0975 & 0.4130 & 0.4105 & 0.3731 & 0.3745 \\ 
           & & VGG-16 & 0.4833 & 0.7189 & 0.4669 & \textbf{0.7162} & 0.5227 & 0.5227 & 0.6355 & 0.6396 & 0.6309 & 0.4675 \\ 
       & NS & ResNet-20 & 0.3952 & \textbf{0.9399} & 0.3702 & 0.9065 & 0.3702 & 0.1316 & 0.4361 & 0.4344 & 0.4103 & 0.3563 \\ 
        &  & VGG-16 & 0.5283 & 0.7711 & 0.4933 & \textbf{0.7820} & 0.5557 & 0.5557 & 0.6594 & 0.6599 & 0.6564 & 0.5441 \\ \hline
        JSMA & GF & ResNet-20 & \textbf{0.5750} & 0.4913 & 0.5207 & 0.4753 & 0.5207 & 0.4620 & 0.4972 & 0.4988 & 0.4844 & 0.4369 \\ 
         &  & VGG-16 & 0.6206 & 1.0161 & 0.6604 & \textbf{1.0261} & 0.5998 & 0.5998 & 0.7440 & 0.7596 & 0.7319 & 0.6443 \\ 
      
         & NAI & ResNet-20 & 0.5546 & \textbf{0.9353} & 0.4947 & 0.7868 & 0.4947 & 0.5077 & 0.6502 & 0.6529 & 0.6346 & 0.538 \\        
         & & VGG-16 & 0.5722 & 0.9640 & 0.5983 & \textbf{0.9792} & 0.5604 & 0.5604 & 0.7071 & 0.7312 & 0.6898 & 0.5791 \\ 

         & NS & ResNet-20 & 0.6210 & \textbf{1.0185} & 0.5365 & 0.9238 & 0.5365 & 0.5749 & 0.7206 & 0.7222 & 0.7103 & 0.6048 \\ 
        
         &  & VGG-16 & 0.6476 & \textbf{1.0361} & 0.6892 & 0.9441 & 0.6238 & 0.6238 & 0.7594 & 0.7717 & 0.7531 & 0.6403 \\ \hline
        \multicolumn{13}{|c|}{FashionMNIST}\\
\hline
        FGSM & GF & ResNet-20 & 0.3824 & 0.4773 & 0.3919 & \textbf{1.0087} & 0.3919 & 0.3919 & 0.5243 & 0.5389 & 0.4531 & 0.3737 \\ 
           &  & VGG-16 & 0.3663 & \textbf{0.6201} & 0.3181 & 0.5601 & 0.3127 & 0.3127 & 0.2787 & 0.3444 & 0.2484 & 0.3156 \\ 
          & NAI & ResNet-20 & 0.3688 & 0.4947 & 0.3983 & \textbf{1.0166} & 0.3983 & 0.3983 & 0.5316 & 0.5452 & 0.4648 & 0.3705 \\ 
           & & VGG-16 & \textbf{0.4206} & 0.3819 & 0.4019 & 0.3395 & 0.3910 & 0.3910 & 0.4150 & 0.3899 & 0.3857 & 0.3703 \\ 
          & NS & ResNet-20 & 0.4093 & 0.5991 & 0.4276 & \textbf{0.6446} & 0.4276 & 0.4276 & 0.5594 & 0.5692 & 0.5109 & 0.3929 \\ 
    &  & VGG-16 & \textbf{0.4238} & 0.3632 & 0.3909 & 0.4181 & 0.3796 & 0.3796 & 0.4051 & 0.4139 & 0.3673 & 0.3858 \\ \hline

        JSMA & GF & ResNet-20 & 0.5179 & 0.5514 & 0.5198 & \textbf{0.9937} & 0.5198 & 0.5020 & 0.7069 & 0.7084 & 0.7009 & 0.5192 \\ 
      
         &  & VGG-16 & 0.4997 & 0.6709 & 0.6081 & \textbf{0.8973} & 0.5257 & 0.4873 & 0.7122 & 0.7227 & 0.7062 & 0.4799 \\ 
       
         & NAI & ResNet-20 & 0.5701 & 0.6201 & 0.5567 & \textbf{0.8206} & 0.5567 & 0.5342 & 0.7296 & 0.7295 & 0.7288 & 0.5469 \\ 
      
         &  & VGG-16 & 0.5672 & 0.6938 & 0.6089 & 0.6119 & 0.6247 & 0.6129 & 0.7708 & \textbf{0.7761} & 0.7650 & 0.6130 \\ 
       
         & NS & ResNet-20 & 0.6205 & \textbf{0.7101} & 0.6104 & 0.5220 & 0.6104 & 0.5839 & 0.6121 & 0.5605 & 0.5659 & 0.5431 \\ 
        
         & & VGG-16 & \textbf{0.5778} & 0.5123 & 0.4193 & 0.4206 & 0.5337 & 0.5269 & 0.5754 & 0.5703 & 0.5701 & 0.5298 \\ \hline
    \end{tabular}
    }
    }
        \caption{Diversity of the test set prioritized by LBT and baseline approaches from adversarial dataset generated from adversarial attack FGSM and JSMA on MUTs trained on MNIST and MNIST}
        \label{tab:diversity}
\end{table*}

\begin{dashedbox}
Although our proposed method is less effective than DSA and KDE in ensuring diversity among the selected inputs, it outperforms most other baseline methods.
\end{dashedbox}
\subsection{RQ4: Retraining the MUT}
We checked if the prioritized adversarial inputs can help retraining the model for improved robustness and accuracy. To check this, we randomly sampled 10\% for evaluation for each generated adversarial dataset, using it to assess the retraining effectiveness of the proposed method compared to baseline approaches. The remaining 90\% serves as adversarial input for prioritization.
We retrained MUTs on this prioritized set and use the evaluation set as a validation set to evaluate the updated accuracy of the MUTs on the evaluation set. 
\textbf{Table~\ref{tab:retrain_acc}} shows the change of the models' accuracy on the evaluation dataset following retraining with the prioritized adversarial subset. 

Overall, our proposed method achieves better accuracy improvement for MUTs on the evaluation test set after retraining with the prioritized subset derived from the proposed method compared to other baselines with some exceptions where either the retraining caused a decrement of accuracy or the improvement is lower than other baselines. However, even when the improvement is not the highest, it is marginally lower than the baseline, which achieved the highest improvement and is an improvement over the original accuracy of the MUT on the corresponding evaluation test set.

A notable finding is that the white-box baseline methods such as KDE, NAC, NBC have shown better accuracy improvement on the evaluation test set in comparison to black-box methods. In addition, in some cases, even random selection can achieve better improvement than any other methods.


\begin{table*}
    \centering
    
      \raisebox{0.05\height}{ 
        \scalebox{0.90}[0.90]{ 
    \begin{tabular}{|c|c|c|c|c|c|c|c|c|c|c|c|c|}
    \hline
                \textbf{Attack} & 
\textbf{BM-MO} & 
\textbf{MUT} & 
\textbf{LBT} & 
\textbf{KDE} & 
\textbf{DR} & 
\textbf{DSA} &
\textbf{NAC} & 
\textbf{NBC} & 
\textbf{DeepG.} & 
\textbf{Maxp} & 
\textbf{PE} & 
\textbf{Rand.} \\ \hline
        \multicolumn{13}{|c|}{MNIST}\\ \hline
        FGSM & GF & ResNet-20 & \textbf{+0.236} & +0.071 & +0.202 & +0.023 & +0.208 & +0.010 & +0.236 & +0.223 & +0.192 & +0.195 \\ 
        &  & VGG-16 & -0.097 & +0.217 & +0.013 & -0.007 & \textbf{+0.225} & +0.202 & -0.049 & -0.025 & -0.045 & +0.189 \\ 
        & NAI & ResNet-20 & \textbf{+0.156} & +0.025 & +0.137 & +0.012 & +0.137 & +0.008 & +0.149 & +0.139 & +0.107 & +0.076 \\ 
          &  & VGG-16 & -0.212 & +0.240 & +0.152 & -0.049 & +0.279 & \textbf{+0.286} & +0.025 & +0.013 & -0.009 & +0.193 \\ 
        & NS & ResNet-20 & \textbf{+0.238} & +0.056 & +0.175 & +0.018 & +0.166 & +0.008 & +0.212 & +0.220 & +0.195 & +0.163 \\ 
         &  & VGG-16 & -0.276 & +0.219 & +0.083 & +0.005 & +0.239 & \textbf{+0.257} & -0.028 & +0.005 & -0.014 & +0.143 \\ \hline
        JSMA & GF & ResNet-20 & +0.109 & +0.041 & +0.120 & -0.163 & \textbf{+0.125} & +0.074 & +0.016 & -0.004 & -0.066 & +0.008\\ 
         &  & VGG-16 & \textbf{+0.241} & +0.232 & +0.202 & +0.207 & +0.221 & +0.230 & +0.085 & +0.106 & +0.062 & +0.238 \\
         & NAI & ResNet-20 & \textbf{+0.169} & +0.114 & +0.166 & +0.025 & +0.168 & +0.098 & +0.042 & +0.083 & +0.011 & +0.137 \\ 
         & & VGG-16 & +0.186 & \textbf{+0.228} & +0.149 & +0.166 & +0.184 & +0.200 & +0.020 & +0.054 & +0.044 & +0.173 \\ 
         & NS & ResNet-20 & \textbf{+0.073} & -0.092 & +0.035 & -0.434 & +0.052 & -0.013 & -0.143 & -0.106 & -0.236 & -0.019 \\ 
        & & VGG-16 & +0.248 & \textbf{+0.253} & +0.236 & +0.238 & +0.239 & +0.234 & +0.202 & +0.214 & +0.179 & +0.251 \\ \hline
                \multicolumn{13}{|c|}{FashionMNIST}\\ \hline
        FGSM & GF & ResNet-20 & +0.139 & +0.030 & +0.136 & \textbf{+0.159} & +0.140 & +0.134 & +0.145 & +0.147 & +0.128 & +0.126 \\ 
        & & VGG-16 & \textbf{+0.122} & +0.078 & +0.112 & +0.020 & \textbf{+0.122} & +0.118 & -0.041 & -0.012 & -0.035 & +0.118 \\ 
         & NAI & ResNet-20 & \textbf{+0.152} & +0.032 & +0.147 & +0.015 & +0.147 & +0.136 & +0.146 & \textbf{+0.152} & +0.133 & +0.141 \\ 
         & & VGG-16 & +0.088 & \textbf{+0.102} & +0.080 & -0.233 & +0.094 & \textbf{+0.102} & -0.318 & -0.154 & -0.383 & +0.087 \\ 
         & NS & ResNet-20 & \textbf{+0.110} & +0.026 & +0.100 & +0.107 & +0.102 & +0.105 & +0.102 & +0.099 & +0.091 & +0.107 \\ 
         & & VGG-16 & +0.099 & +0.090 & +0.099 & -0.099 & +0.107 & \textbf{+0.109} & -0.189 & -0.122 & -0.255 & +0.095 \\ \hline
        JSMA & GF & ResNet-20 & +0.058 & +0.043 & +0.070 & +0.048 & \textbf{+0.071} & +0.062 & +0.047 & +0.039 & +0.058 & +0.067 \\ 
         &  & VGG-16 & +0.040 & \textbf{+0.060} & +0.014 & +0.012 & +0.059 & +0.044 & +0.046 & +0.034 & +0.034 & +0.049 \\ 
         & NAI & ResNet-20 & \textbf{+0.059} & +0.057 & +0.057 & +0.052 & +0.039 & +0.038 & +0.056 & +0.045 & +0.055 & +0.045 \\ 
         &  & VGG-16 & +0.062 & +0.010 & +0.006 & +0.018 & +0.082 & +0.073 & +0.031 & +0.067 & +0.031 & \textbf{+0.086} \\ 
         & NS & ResNet-20 & \textbf{+0.068} & +0.058 & +0.066 & +0.034 & 0.050 & +0.060 & +0.062 & +0.065 & +0.062 & +0.057 \\ 
         &  & VGG-16 & +0.048 & \textbf{+0.097} & +0.031 & +0.024 & +0.078 & +0.072 & +0.042 & +0.059 & +0.041 & +0.074 \\ \hline
    \end{tabular}
    }
    }
     \caption{Accuracy improvement on evaluation set by retraining the MUTs with the test set prioritized by LBT and baseline approaches from adversarial dataset.}
    \label{tab:retrain_acc}
\end{table*}
\begin{dashedbox}
Our proposed method more effectively prioritizes adversarial inputs, resulting in consistently higher MUT accuracy improvement on the evaluation test set compared to baseline methods.
\end{dashedbox}



\section{Discussion}
We demonstrate that our proposed method offers superior adversarial test case prioritization compared to existing baselines in selecting test inputs with high fault-revealing potential and in the order in which the prioritized test cases are executed. The second-best approaches for these metrics, FDR and APFD, vary across different models and datasets, showing no consistent pattern. 

The results show that white-box methods such as KDE and DSA performed better at ensuring diversity in selected adversarial test inputs. One of the reasons is that these prioritization methods pick inputs based on surprise adequacy, meaning how inputs are different from the training and already prioritized inputs. However, they are ineffective across different MUT and adversarial datasets at detecting faults and ensuring that the permutation of the selected inputs can do it faster. Moreover, these methods can not be applied for third-party pre-trained DNN models or legacy systems, where access to internal model details is usually unavailable.


It is also evident that confidence-based black-box metrics can achieve high FDR in certain cases. However, the FDR of the subset selected by the three confidence-based metrics is inconsistent, occasionally dropping to average values. Several factors contribute to this. \citet{dang2024test} demonstrated that certain limitations arise when adapting confidence-based test prioritization approaches. Firstly, confidence-based approaches, treating the model as a black box and relying solely on prediction probabilities, overlook the transparency and interpretability inherent in models, thus leading to sub-optimal prioritization. Secondly, these methods disregard the attribute features of classical ML test datasets, which play a crucial role in mapping tests and indirectly reflecting the proximity of samples to the model's decision boundary. Thirdly, these methods may prove ineffective in some specific situations. For example, in binary classification models, such approaches primarily rely on the model's prediction probability, resulting in a single dimension for prioritization. Tests with probabilities near 0.5 are consistently prioritized, regardless of the method. Moreover, these methods, like white-box ones, become inapplicable in a fully black-box scenario without softmax access.

Our proposed approach draws inspiration from both white-box and black-box methods. We treat the MUT as a fully black-box system, meaning we can only observe its outputs. However, we introduce mutations in the behavioral model, a white-box model with a known structure. While the white-box model used in our experiments may be significantly simpler than the MUTs, this aligns with the core idea of LBT—using a simplified model to represent the MUT~\cite{sharma2021mlcheck,papadopoulos2015black,sharma2022property}. Unlike traditional LBT methods that require the behavioral model to be encodable for SMT solver-based test generation, which often leads to ineffective or invalid test cases, our approach removes this constraint. Instead, we focus on prioritizing test generation strategies to maximize fault detection. 

A key implication of this research is that our method is not limited to DNN model testing but can be applied to any system that a DNN can represent. Furthermore, this technique is adaptable to different model types (which can represent the SUT) by employing mutation operators specific to their structures. Exploring its applicability to a broader range of model types and SUTs could be a promising future research direction.


\section{Limitation}
One potential limitation of this approach is that the behavioral models trained for black-box attacks may not achieve state-of-the-art accuracy. This could lead to a high mutation score even for benign test cases when mutated, as the adversary typically has access to only a limited number of samples. However, the primary objective is not to attain optimal accuracy but to approximate the oracle’s decision boundaries.

Another limitation is the computational cost of generating multiple mutants of the behavioral model. A high mutation rate can lead to mutants that diverge significantly from the original model, resulting in high mutation scores even for benign inputs. Conversely, a low mutation rate may produce mutants that closely resemble the original model, leading to low mutation scores even for adversarial inputs that could cause the MUT to misclassify. For instance, in the case of a LeNet-5-based behavioral model, mutants generated using the model-level mutation operator NAI with a low mutation rate produced identical outputs to the original model for all adversarial inputs. Moreover, some mutation operators may fail to be effective even at high mutation rates, generating mutants that remain behaviorally similar to the original model. 
For example, the WS mutation operator consistently generated mutants with identical outputs to the original model for all inputs.

In addition, regression tasks require solving a challenge, as most test case prioritization methods focus on prioritizing test inputs for classification models. In regression models, particularly those based on neural networks, there is no direct analog to the softmax function, which inherently provides a way to calculate mutation score for the adversarial inputs based on classes.  
\vspace{10pt}
\textbf{\textit{Threats to Validity.}} We addressed internal, construct, and external validity to ensure the robustness and reliability of our findings. 

\begin{itemize}
    \item To mitigate \textbf{internal validity} concerns, we generated adversarial datasets using different attack techniques and ensured that the MUTs maintained consistent accuracy levels on adversarial and benign inputs.

    \item For \textbf{construct validity}, we used two state-of-the-art adversarial attack techniques to generate adversarial test sets for each of the MUTs, allowing us to accurately measure the effectiveness of the proposed LBT method using four different metrics.

    \item To enhance \textbf{external validity} and support generalizability, we evaluated the method against different combinations of widely recognized models, datasets, and attack types, included various testing budgets in our experiments, and benchmarked our results against state-of-the-art DNN test selection baselines.

    \item Moreover, to ensure conclusive and reliable experiments, we selected baselines, baseline parameters, MUT architectures, attack types, and datasets based on established research works.
\end{itemize}

\section{Conclusion}
In this paper, we introduced an LBT-based method to prioritize adversarial test cases with a high likelihood of being misclassified. Using widely recognized adversarial attack techniques with manually configured parameters, we generated adversarial inputs from two popular datasets. We compared the adversarial test sets prioritized by our proposed approach with other leading prioritization techniques using evaluation metrics used in existing research that assess their ability to select fault-detecting inputs while considering factors such as effective permutation, diversity, and guidance on retraining the MUTs. Results show that our LBT-based method effectively prioritizes high-FDR adversarial inputs across models, attacks, and datasets while optimizing test execution for faster detection. Furthermore, our findings suggest that our method provides valuable guidance for model retraining.

As mentioned in the discussion section, this approach can be applied to even more complex models or systems, as long as the system can be expressed as a deep learning model, due to the nature of the LBT-based approach. So, the applicability of the approach is not confined to DNN testing alone.
For future work, we plan to evaluate the effectiveness of the proposed method in prioritizing test cases for assessing systems that integrate both data-driven and code-based characteristics, particularly non-deterministic systems where conventional testing methods may fall short.



\bibliographystyle{abbrvnat}
\bibliography{references}  

\begin{thebibliography}{56}
\providecommand{\natexlab}[1]{#1}
\providecommand{\url}[1]{\texttt{#1}}
\expandafter\ifx\csname urlstyle\endcsname\relax
  \providecommand{\doi}[1]{doi: #1}\else
  \providecommand{\doi}{doi: \begingroup \urlstyle{rm}\Url}\fi

\bibitem[Abadi et~al.(2015)Abadi, Agarwal, Barham, Brevdo, Chen, Citro,
  Corrado, Davis, Dean, Devin, Ghemawat, Goodfellow, Harp, Irving, Isard, Jia,
  Jozefowicz, Kaiser, Kudlur, Levenberg, Man\'{e}, Monga, Moore, Murray, Olah,
  Schuster, Shlens, Steiner, Sutskever, Talwar, Tucker, Vanhoucke, Vasudevan,
  Vi\'{e}gas, Vinyals, Warden, Wattenberg, Wicke, Yu, and
  Zheng]{tensorflow2015-whitepaper}
M.~Abadi, A.~Agarwal, P.~Barham, E.~Brevdo, Z.~Chen, C.~Citro, G.~S. Corrado,
  A.~Davis, J.~Dean, M.~Devin, S.~Ghemawat, I.~Goodfellow, A.~Harp, G.~Irving,
  M.~Isard, Y.~Jia, R.~Jozefowicz, L.~Kaiser, M.~Kudlur, J.~Levenberg,
  D.~Man\'{e}, R.~Monga, S.~Moore, D.~Murray, C.~Olah, M.~Schuster, J.~Shlens,
  B.~Steiner, I.~Sutskever, K.~Talwar, P.~Tucker, V.~Vanhoucke, V.~Vasudevan,
  F.~Vi\'{e}gas, O.~Vinyals, P.~Warden, M.~Wattenberg, M.~Wicke, Y.~Yu, and
  X.~Zheng.
\newblock {TensorFlow}: Large-scale machine learning on heterogeneous systems,
  2015.
\newblock URL \url{https://www.tensorflow.org/}.
\newblock Software available from tensorflow.org.

\bibitem[Aghababaeyan et~al.(2024)Aghababaeyan, Abdellatif, Dadkhah, and
  Briand]{aghababaeyan2024deepgd}
Z.~Aghababaeyan, M.~Abdellatif, M.~Dadkhah, and L.~Briand.
\newblock Deepgd: A multi-objective black-box test selection approach for deep
  neural networks.
\newblock \emph{ACM Transactions on Software Engineering and Methodology},
  33\penalty0 (6):\penalty0 1--29, 2024.

\bibitem[Arrieta(2022)]{arrieta2022multi}
A.~Arrieta.
\newblock Multi-objective metamorphic follow-up test case selection for deep
  learning systems.
\newblock In \emph{Proceedings of the Genetic and Evolutionary Computation
  Conference}, pages 1327--1335, 2022.

\bibitem[Bojarski et~al.(2016)Bojarski, Del~Testa, Dworakowski, Firner, Flepp,
  Goyal, Jackel, Monfort, Muller, Zhang, et~al.]{bojarski2016end}
M.~Bojarski, D.~Del~Testa, D.~Dworakowski, B.~Firner, B.~Flepp, P.~Goyal, L.~D.
  Jackel, M.~Monfort, U.~Muller, J.~Zhang, et~al.
\newblock End to end learning for self-driving cars.
\newblock \emph{arXiv preprint arXiv:1604.07316}, 2016.

\bibitem[Briand et~al.(2009)Briand, Labiche, Bawar, and Spido]{briand2009using}
L.~C. Briand, Y.~Labiche, Z.~Bawar, and N.~T. Spido.
\newblock Using machine learning to refine category-partition test
  specifications and test suites.
\newblock \emph{Information and Software Technology}, 51\penalty0
  (11):\penalty0 1551--1564, 2009.

\bibitem[Budd and Angluin(1982)]{budd1982two}
T.~A. Budd and D.~Angluin.
\newblock Two notions of correctness and their relation to testing.
\newblock \emph{Acta informatica}, 18:\penalty0 31--45, 1982.

\bibitem[Carlini and Wagner(2017)]{carlini2017adversarial}
N.~Carlini and D.~Wagner.
\newblock Adversarial examples are not easily detected: Bypassing ten detection
  methods.
\newblock In \emph{Proceedings of the 10th ACM workshop on artificial
  intelligence and security}, pages 3--14, 2017.

\bibitem[Chen et~al.(2005)Chen, Leung, and Mak]{chen2005adaptive}
T.~Y. Chen, H.~Leung, and I.~K. Mak.
\newblock Adaptive random testing.
\newblock In \emph{Advances in Computer Science-ASIAN 2004. Higher-Level
  Decision Making: 9th Asian Computing Science Conference. Dedicated to
  Jean-Louis Lassez on the Occasion of His 5th Birthday. Chiang Mai, Thailand,
  December 8-10, 2004. Proceedings 9}, pages 320--329. Springer, 2005.

\bibitem[Chollet(2021)]{chollet2021deep}
F.~Chollet.
\newblock \emph{Deep learning with Python}.
\newblock Simon and Schuster, 2021.

\bibitem[Dang et~al.(2024)Dang, Li, Papadakis, Klein, Bissyand{\'e}, and
  Le~Traon]{dang2024test}
X.~Dang, Y.~Li, M.~Papadakis, J.~Klein, T.~F. Bissyand{\'e}, and Y.~Le~Traon.
\newblock Test input prioritization for machine learning classifiers.
\newblock \emph{IEEE Transactions on Software Engineering}, 2024.

\bibitem[Davies(2019)]{davies2019tesla}
A.~Davies.
\newblock Tesla’s latest autopilot death looks just like a prior crash.
\newblock \emph{WIRED Magazine}, 2019.

\bibitem[Deng(2012)]{deng2012mnist}
L.~Deng.
\newblock The mnist database of handwritten digit images for machine learning
  research.
\newblock \emph{IEEE Signal Processing Magazine}, 29\penalty0 (6):\penalty0
  141--142, 2012.

\bibitem[Feinman et~al.(2017)Feinman, Curtin, Shintre, and
  Gardner]{feinman2017detecting}
R.~Feinman, R.~R. Curtin, S.~Shintre, and A.~B. Gardner.
\newblock Detecting adversarial samples from artifacts.
\newblock \emph{arXiv preprint arXiv:1703.00410}, 2017.

\bibitem[Feng et~al.(2020)Feng, Shi, Gao, Wan, Fang, and
  Chen]{feng2020deepgini}
Y.~Feng, Q.~Shi, X.~Gao, J.~Wan, C.~Fang, and Z.~Chen.
\newblock Deepgini: prioritizing massive tests to enhance the robustness of
  deep neural networks.
\newblock In \emph{Proceedings of the 29th ACM SIGSOFT International Symposium
  on Software Testing and Analysis}, pages 177--188, 2020.

\bibitem[Fraser and Walkinshaw(2012)]{fraser2012behaviourally}
G.~Fraser and N.~Walkinshaw.
\newblock Behaviourally adequate software testing.
\newblock In \emph{2012 IEEE Fifth Intl. conf. on Software Testing,
  Verification and Validation}, 2012.

\bibitem[Fraser and Walkinshaw(2015)]{fraser2015assessing}
G.~Fraser and N.~Walkinshaw.
\newblock Assessing and generating test sets in terms of behavioural adequacy.
\newblock \emph{Software Testing, Verification and Reliability}, 2015.

\bibitem[Gao et~al.(2022)Gao, Feng, Yin, Liu, Chen, and Xu]{gao2022adaptive}
X.~Gao, Y.~Feng, Y.~Yin, Z.~Liu, Z.~Chen, and B.~Xu.
\newblock Adaptive test selection for deep neural networks.
\newblock In \emph{Proceedings of the 44th International Conference on Software
  Engineering}, pages 73--85, 2022.

\bibitem[Gerasimou et~al.(2020)Gerasimou, Eniser, Sen, and
  Cakan]{gerasimou2020importance}
S.~Gerasimou, H.~F. Eniser, A.~Sen, and A.~Cakan.
\newblock Importance-driven deep learning system testing.
\newblock In \emph{Proceedings of the ACM/IEEE 42nd International Conference on
  Software Engineering}, pages 702--713, 2020.

\bibitem[Goodfellow et~al.(2014)Goodfellow, Shlens, and
  Szegedy]{goodfellow2014explaining}
I.~J. Goodfellow, J.~Shlens, and C.~Szegedy.
\newblock Explaining and harnessing adversarial examples.
\newblock \emph{arXiv preprint arXiv:1412.6572}, 2014.

\bibitem[He et~al.(2016)He, Zhang, Ren, and Sun]{he2016deep}
K.~He, X.~Zhang, S.~Ren, and J.~Sun.
\newblock Deep residual learning for image recognition.
\newblock In \emph{Proceedings of the IEEE conference on computer vision and
  pattern recognition}, pages 770--778, 2016.

\bibitem[Jinliang(2019)]{jinliang2019cifar10}
N.~Jinliang.
\newblock Cifar10 image classification based on resnet.
\newblock \emph{Системный анализ в проектировании
  и управлении}, 23\penalty0 (1):\penalty0 412--415, 2019.

\bibitem[Kim et~al.(2019)Kim, Feldt, and Yoo]{kim2019guiding}
J.~Kim, R.~Feldt, and S.~Yoo.
\newblock Guiding deep learning system testing using surprise adequacy.
\newblock In \emph{2019 IEEE/ACM 41st International Conference on Software
  Engineering (ICSE)}, pages 1039--1049. IEEE, 2019.

\bibitem[Li et~al.(2019)Li, Ma, Xu, Cao, Xu, and L{\"u}]{li2019boosting}
Z.~Li, X.~Ma, C.~Xu, C.~Cao, J.~Xu, and J.~L{\"u}.
\newblock Boosting operational dnn testing efficiency through conditioning.
\newblock In \emph{Proceedings of the 2019 27th ACM Joint Meeting on European
  Software Engineering Conference and Symposium on the Foundations of Software
  Engineering}, pages 499--509, 2019.

\bibitem[Ma et~al.(2018{\natexlab{a}})Ma, Juefei-Xu, Zhang, Sun, Xue, Li, Chen,
  Su, Li, Liu, et~al.]{ma2018deepgauge}
L.~Ma, F.~Juefei-Xu, F.~Zhang, J.~Sun, M.~Xue, B.~Li, C.~Chen, T.~Su, L.~Li,
  Y.~Liu, et~al.
\newblock Deepgauge: Multi-granularity testing criteria for deep learning
  systems.
\newblock In \emph{Proceedings of the 33rd ACM/IEEE international conference on
  automated software engineering}, pages 120--131, 2018{\natexlab{a}}.

\bibitem[Ma et~al.(2018{\natexlab{b}})Ma, Zhang, Sun, Xue, Li, Juefei-Xu, Xie,
  Li, Liu, Zhao, et~al.]{ma2018deepmutation}
L.~Ma, F.~Zhang, J.~Sun, M.~Xue, B.~Li, F.~Juefei-Xu, C.~Xie, L.~Li, Y.~Liu,
  J.~Zhao, et~al.
\newblock Deepmutation: Mutation testing of deep learning systems.
\newblock In \emph{2018 IEEE 29th international symposium on software
  reliability engineering (ISSRE)}, pages 100--111. IEEE, 2018{\natexlab{b}}.

\bibitem[Ma et~al.(2021)Ma, Papadakis, Tsakmalis, Cordy, and Traon]{ma2021test}
W.~Ma, M.~Papadakis, A.~Tsakmalis, M.~Cordy, and Y.~L. Traon.
\newblock Test selection for deep learning systems.
\newblock \emph{ACM Transactions on Software Engineering and Methodology
  (TOSEM)}, 30\penalty0 (2):\penalty0 1--22, 2021.

\bibitem[Meinke and Niu(2010)]{meinke2010learning}
K.~Meinke and F.~Niu.
\newblock A learning-based approach to unit testing of numerical software.
\newblock In \emph{IFIP International Conference on Testing Software and
  Systems}, pages 221--235. Springer, 2010.

\bibitem[Nicolae et~al.(2018)Nicolae, Sinn, Tran, Buesser, Rawat, Wistuba,
  Zantedeschi, Baracaldo, Chen, Ludwig, Molloy, and Edwards]{art2018}
M.-I. Nicolae, M.~Sinn, M.~N. Tran, B.~Buesser, A.~Rawat, M.~Wistuba,
  V.~Zantedeschi, N.~Baracaldo, B.~Chen, H.~Ludwig, I.~Molloy, and B.~Edwards.
\newblock Adversarial robustness toolbox v1.2.0.
\newblock \emph{CoRR}, 1807.01069, 2018.
\newblock URL \url{https://arxiv.org/pdf/1807.01069}.

\bibitem[Papadopoulos and Walkinshaw(2015)]{papadopoulos2015black}
P.~Papadopoulos and N.~Walkinshaw.
\newblock Black-box test generation from inferred models.
\newblock In \emph{2015 IEEE/ACM 4th International Workshop on Realizing
  Artificial Intelligence Synergies in Software Engineering}, pages 19--24.
  IEEE, 2015.

\bibitem[Papernot et~al.(2016)Papernot, McDaniel, Jha, Fredrikson, Celik, and
  Swami]{papernot2016limitations}
N.~Papernot, P.~McDaniel, S.~Jha, M.~Fredrikson, Z.~B. Celik, and A.~Swami.
\newblock The limitations of deep learning in adversarial settings.
\newblock In \emph{2016 IEEE European symposium on security and privacy
  (EuroS\&P)}, pages 372--387. IEEE, 2016.

\bibitem[Pei et~al.(2017)Pei, Cao, Yang, and Jana]{pei2017deepxplore}
K.~Pei, Y.~Cao, J.~Yang, and S.~Jana.
\newblock Deepxplore: Automated whitebox testing of deep learning systems.
\newblock In \emph{proceedings of the 26th Symposium on Operating Systems
  Principles}, pages 1--18, 2017.

\bibitem[Sharma and Wehrheim(2020)]{sharma2020higher}
A.~Sharma and H.~Wehrheim.
\newblock Higher income, larger loan? monotonicity testing of machine learning
  models.
\newblock In \emph{Proceedings of the 29th ACM SIGSOFT International Symposium
  on Software Testing and Analysis}, pages 200--210, 2020.

\bibitem[Sharma et~al.(2021)Sharma, Demir, Ngomo, and
  Wehrheim]{sharma2021mlcheck}
A.~Sharma, C.~Demir, A.-C.~N. Ngomo, and H.~Wehrheim.
\newblock Mlcheck--property-driven testing of machine learning classifiers.
\newblock In \emph{2021 20th IEEE International Conference on Machine Learning
  and Applications (ICMLA)}, pages 738--745. IEEE, 2021.

\bibitem[Sharma et~al.(2022)Sharma, Melnikov, H{\"u}llermeier, and
  Wehrheim]{sharma2022property}
A.~Sharma, V.~Melnikov, E.~H{\"u}llermeier, and H.~Wehrheim.
\newblock Property-driven testing of black-box functions.
\newblock In \emph{Proceedings of the IEEE/ACM 10th Intl. conf. on Formal
  Methods in Software Engineering}, pages 113--123, 2022.

\bibitem[Shen et~al.(2020)Shen, Li, Chen, Han, Zhou, and Xu]{shen2020multiple}
W.~Shen, Y.~Li, L.~Chen, Y.~Han, Y.~Zhou, and B.~Xu.
\newblock Multiple-boundary clustering and prioritization to promote neural
  network retraining.
\newblock In \emph{Proceedings of the 35th IEEE/ACM International Conference on
  Automated Software Engineering}, pages 410--422, 2020.

\bibitem[Shi et~al.(2021)Shi, Yin, Zheng, and Li]{shi2021empirical}
Y.~Shi, B.~Yin, Z.~Zheng, and T.~Li.
\newblock An empirical study on test case prioritization metrics for deep
  neural networks.
\newblock In \emph{2021 IEEE 21st International Conference on Software Quality,
  Reliability and Security (QRS)}, pages 157--166. IEEE, 2021.

\bibitem[Simonyan and Zisserman(2014)]{simonyan2014very}
K.~Simonyan and A.~Zisserman.
\newblock Very deep convolutional networks for large-scale image recognition.
\newblock \emph{arXiv preprint arXiv:1409.1556}, 2014.

\bibitem[Sun et~al.(2018)Sun, Wang, and Dai]{sun2018convolutional}
L.~Sun, Y.~Wang, and L.~Dai.
\newblock Convolutional neural network protection method of lenet-5-like
  structure.
\newblock In \emph{Proceedings of the 2018 2nd International Conference on
  Computer Science and Artificial Intelligence}, pages 77--80, 2018.

\bibitem[Tian et~al.(2018)Tian, Pei, Jana, and Ray]{tian2018deeptest}
Y.~Tian, K.~Pei, S.~Jana, and B.~Ray.
\newblock Deeptest: Automated testing of deep-neural-network-driven autonomous
  cars.
\newblock In \emph{Proceedings of the 40th international conference on software
  engineering}, pages 303--314, 2018.

\bibitem[Wald(2004)]{wald2004sequential}
A.~Wald.
\newblock \emph{Sequential analysis}.
\newblock Courier Corporation, 2004.

\bibitem[Walkinshaw and Fraser(2017)]{walkinshaw2017uncertainty}
N.~Walkinshaw and G.~Fraser.
\newblock Uncertainty-driven black-box test data generation.
\newblock In \emph{2017 IEEE International Conference on Software Testing,
  Verification and Validation (ICST)}. IEEE, 2017.

\bibitem[Walkinshaw et~al.(2010)Walkinshaw, Bogdanov, Derrick, and
  Paris]{walkinshaw2010increasing}
N.~Walkinshaw, K.~Bogdanov, J.~Derrick, and J.~Paris.
\newblock Increasing functional coverage by inductive testing: A case study.
\newblock In \emph{Testing Software and Systems: 22nd IFIP WG 6.1 International
  Conference, ICTSS 2010, Natal, Brazil, November 8-10, 2010. Proceedings 22},
  pages 126--141. Springer, 2010.

\bibitem[Wang et~al.(2021)Wang, You, Chen, Zhang, Dong, and
  Zhang]{wang2021prioritizing}
Z.~Wang, H.~You, J.~Chen, Y.~Zhang, X.~Dong, and W.~Zhang.
\newblock Prioritizing test inputs for deep neural networks via mutation
  analysis.
\newblock In \emph{2021 IEEE/ACM 43rd International Conference on Software
  Engineering (ICSE)}, pages 397--409. IEEE, 2021.

\bibitem[Weiss and Tonella(2022{\natexlab{a}})]{10.1145/3533767.3534375}
M.~Weiss and P.~Tonella.
\newblock Simple techniques work surprisingly well for neural network test
  prioritization and active learning (replicability study).
\newblock In \emph{Proceedings of the 31st ACM SIGSOFT International Symposium
  on Software Testing and Analysis}, ISSTA 2022, page 139–150, New York, NY,
  USA, 2022{\natexlab{a}}. Association for Computing Machinery.
\newblock ISBN 9781450393799.
\newblock \doi{10.1145/3533767.3534375}.
\newblock URL \url{https://doi.org/10.1145/3533767.3534375}.

\bibitem[Weiss and Tonella(2022{\natexlab{b}})]{weiss2022simple}
M.~Weiss and P.~Tonella.
\newblock Simple techniques work surprisingly well for neural network test
  prioritization and active learning (replicability study).
\newblock In \emph{Proceedings of the 31st ACM SIGSOFT International Symposium
  on Software Testing and Analysis}, pages 139--150, 2022{\natexlab{b}}.

\bibitem[Weyuker(1983)]{weyuker1983assessing}
E.~J. Weyuker.
\newblock Assessing test data adequacy through program inference.
\newblock \emph{ACM Transactions on Programming Languages and Systems
  (TOPLAS)}, 1983.

\bibitem[Wong et~al.(1995)Wong, Horgan, London, and Mathur]{wong1995effect}
W.~E. Wong, J.~R. Horgan, S.~London, and A.~P. Mathur.
\newblock Effect of test set minimization on fault detection effectiveness.
\newblock In \emph{Proceedings of the 17th international conference on Software
  engineering}, pages 41--50, 1995.

\bibitem[Xiao et~al.(2017)Xiao, Rasul, and Vollgraf]{xiao2017fashion}
H.~Xiao, K.~Rasul, and R.~Vollgraf.
\newblock Fashion-mnist: a novel image dataset for benchmarking machine
  learning algorithms.
\newblock \emph{arXiv preprint arXiv:1708.07747}, 2017.

\bibitem[Xie et~al.(2022)Xie, Li, Wang, Ma, Guo, Juefei-Xu, and
  Liu]{xie2022npc}
X.~Xie, T.~Li, J.~Wang, L.~Ma, Q.~Guo, F.~Juefei-Xu, and Y.~Liu.
\newblock Npc: N euron p ath c overage via characterizing decision logic of
  deep neural networks.
\newblock \emph{ACM Transactions on Software Engineering and Methodology
  (TOSEM)}, 31\penalty0 (3):\penalty0 1--27, 2022.

\bibitem[Xiong et~al.(2016)Xiong, Droppo, Huang, Seide, Seltzer, Stolcke, Yu,
  and Zweig]{xiong2016achieving}
W.~Xiong, J.~Droppo, X.~Huang, F.~Seide, M.~Seltzer, A.~Stolcke, D.~Yu, and
  G.~Zweig.
\newblock Achieving human parity in conversational speech recognition.
\newblock \emph{arXiv preprint arXiv:1610.05256}, 2016.

\bibitem[Yoo and Harman(2012)]{yoo2012regression}
S.~Yoo and M.~Harman.
\newblock Regression testing minimization, selection and prioritization: a
  survey.
\newblock \emph{Software testing, verification and reliability}, 22\penalty0
  (2):\penalty0 67--120, 2012.

\bibitem[Zhan et~al.(2022)Zhan, Wang, Huang, Xiong, Dou, and
  Chan]{zhan2022comparative}
X.~Zhan, Q.~Wang, K.-h. Huang, H.~Xiong, D.~Dou, and A.~B. Chan.
\newblock A comparative survey of deep active learning.
\newblock \emph{arXiv preprint arXiv:2203.13450}, 2022.

\bibitem[Zhang et~al.(2020)Zhang, Xie, Ma, Du, Hu, Liu, Zhao, and
  Sun]{zhang2020towards}
X.~Zhang, X.~Xie, L.~Ma, X.~Du, Q.~Hu, Y.~Liu, J.~Zhao, and M.~Sun.
\newblock Towards characterizing adversarial defects of deep learning software
  from the lens of uncertainty.
\newblock In \emph{Proceedings of the ACM/IEEE 42nd International Conference on
  Software Engineering}, pages 739--751, 2020.

\bibitem[Zhu and Bento(2017)]{zhu2017generative}
J.-J. Zhu and J.~Bento.
\newblock Generative adversarial active learning.
\newblock \emph{arXiv preprint arXiv:1702.07956}, 2017.

\bibitem[Ziegler(2016)]{ziegler2016google}
C.~Ziegler.
\newblock A google self-driving car caused a crash for the first time.
\newblock \emph{The Verge}, 198, 2016.

\bibitem[Zolfagharian et~al.(2024)Zolfagharian, Abdellatif, Briand, and
  Ramesh]{zolfagharian2024smarla}
A.~Zolfagharian, M.~Abdellatif, L.~C. Briand, and S.~Ramesh.
\newblock Smarla: A safety monitoring approach for deep reinforcement learning
  agents.
\newblock \emph{IEEE Transactions on Software Engineering}, 2024.

\end{thebibliography}
\end{document}